\documentclass{camera}
\usepackage{graphicx}  
\usepackage{gensymb}

\begin{document}

%
\title{Coupled-channels description of the $^{40}$Ca+$^{58,64}$Ni transfer and fusion reactions}

%
\author{G. Scamps$^1$, D. Bourgin$^{2,3}$, K. Hagino$^{1}$, F. Haas$^{2,3}$, \and S. Courtin$^{2,3,4}$ }

%
\organization{

$^1$ Department of Physics, Tohoku University, Sendai 980-8578, Japan

$^2$ IPHC, Universit\'e de Strasbourg, F-67037 Strasbourg, France 

$^3$ CNRS, UMR7178, F-67037 Strasbourg, France

$^4$ USIAS, F-67083 Strasbourg, France

}

\maketitle

\begin{abstract}

Preliminary experimental data for nucleon transfer reactions of 
the $^{40}$Ca+$^{58}$Ni and $^{40}$Ca+$^{64}$Ni systems are analyzed with the coupled-channels 
approach.  It is shown that a simple treatment for the transfer in the coupled-channels method cannot reproduce simultaneously the transfer probabilities and the sub-barrier enhancement of fusion 
cross sections.

\end{abstract}

%


An extra enhancement of fusion cross sections at low energies is often attributed to different 
transfer processes. Several approaches have been employed in order to reproduce this enhancement. 
In our previous analysis \cite{Sca15}, we attempted to reproduce simultaneously the experimental transfer probabilities and the fusion cross sections for the reaction $^{96}$Zr+$^{40}$Ca \cite{Cor11}. The strategy was to adjust the transfer coupling to the experimental transfer data and to compare the fusion cross sections obtained with this approach to the experimental data. A slight underestimation of the fusion cross sections was found at low energies. On the other hand, Esbensen \textit{et al.} \cite{Esb16} showed that it was possible to reproduce perfectly the fusion  cross sections if one takes 
into account in addition the proton transfer channels. 

In this contribution, we test this approach on the $^{40}$Ca+$^{58}$Ni and $^{40}$Ca+$^{64}$Ni systems, for which the fusion cross sections have been measured in ref. \cite{Bou14}. The TDHF theory \cite{Bou16} has shown that the major difference between these two systems is due to the neutron transfer. It is expected that the transfer probability for the $^{40}$Ca+$^{64}$Ni system is by about one order of magnitude higher than that for the  $^{40}$Ca+$^{58}$Ni system. In the present contribution, we shall try to reproduce the neutron and proton transfer channels simultaneously with the enhancement of the fusion cross sections 
at low energies.


The experiment has been recently performed at LNL in inverse kinematics at energies around and below the Coulomb barrier. $^{64}$Ni and $^{58}$Ni beams were delivered by the XTU Tandem accelerator ($E_{\rm lab}$=174, 190 and 210 MeV for $^{64}$Ni, and $E_{\rm lab}$=170 and 201 MeV for $^{58}$Ni) onto a 100 micro g/cm$^2$ strip $^{40}$Ca target deposited on a 15 micro g/cm$^2$ $^{12}$C backing. The $^{40}$Ca target-like ions were detected and identified by the large acceptance ( $\pm$ 5$^{\circ}$) PRISMA magnetic spectrometer around $\theta_{\rm lab}$ = 45$^{\circ}$. Four transfer channels were  identified: +1n, +2n, -1p and -2p, where the + sign means a transfer from the Ni fragment 
to the $^{40}$Ca one.

 
The coupled-channels calculations have been carried out with the program CCFULL \cite{Hag99}, that has been modified in order to compute the neutron and proton transfer probabilities.

\begin{table}[h]
\caption{The values of Woods-Saxon real and imaginary potentials employed in the coupled-channels 
calculations. }
\centering \begin{tabular}{|c|c|c|c|c|c|c|c|}
\hline\hline
System & $V_0$ [MeV]  &  $r_0$ [fm] & $a$ [fm] &  $W_0$ [MeV]  & $r_W$ [fm] &  $a_W$ [fm]   \\
 \hline
   $^{40}$Ca+$^{58}$Ni & 60.41 & 1.17 & 0.66  & 35 & 1.1 & 0.2     \\
   $^{40}$Ca+$^{64}$Ni & 68.70 & 1.16 & 0.66  & 35 & 1.1 & 0.2      \\ 
\hline\hline
\end{tabular}
\label{tab:pot_param}
\end{table}

 The same method as in ref. \cite{Sca15} has been used, with values for the nucleus-nucleus potential given in tab. \ref{tab:pot_param} and with the coupling parameters to the transfer channel in tab. \ref{tab:coupl_param}. This model assumes that the pair transfer is a sum of two processes, that is, a direct pair transfer and a sequential transfer. The sequential transfer coupling is assumed to be equal to the single transfer, i.e. $a_{12}$=$a_{01}$. As in ref. \cite{Bou14}, we take into account the excitation to the first $3^-$ state in $^{40}$Ca at energy 3.736 MeV with a coupling strength of $\beta=0.4$, the $2^+$ state in $^{58}$Ni at 1.454 MeV with $\beta=0.18$, and the $2^+$ state in $^{58}$Ni at  1.346 MeV with $\beta=0.16$.


\begin{table}[h]
\caption{Coupling constants for the transfer channels in the coupled-channels calculations. The parameters $a$ are given in fm, $\beta$ in MeV.fm and $Q$ in MeV. }
\centering \begin{tabular}{|c|c|c|c|c|c|c|c|}
\hline\hline
System & n,p & $a_{01}$  &  $\beta_{01}$  & $Q_{01}$  &  $a_{02}$  & $\beta_{02}$  &  $Q_{02}$    \\
 \hline
   $^{40}$Ca+$^{58}$Ni & n & 1.96 & -6  & -3.85 & 1.06 & -3 & -2.62         \\
   $^{40}$Ca+$^{58}$Ni & p & 1.96 & -16  & -4.909 & 1.06 & -12 & -6.19   \\
   $^{40}$Ca+$^{64}$Ni & n & 1.16 & -37  & -1.29 & 1.06 & -27.39 & 0      \\
   $^{40}$Ca+$^{64}$Ni & p & 1.16 & -30  & -0.874 & 1.06 & -20.39 & 0    \\
\hline\hline
\end{tabular}
\label{tab:coupl_param}
\end{table}

\begin{figure}[!h]
  \centerline{\includegraphics[width= \linewidth]{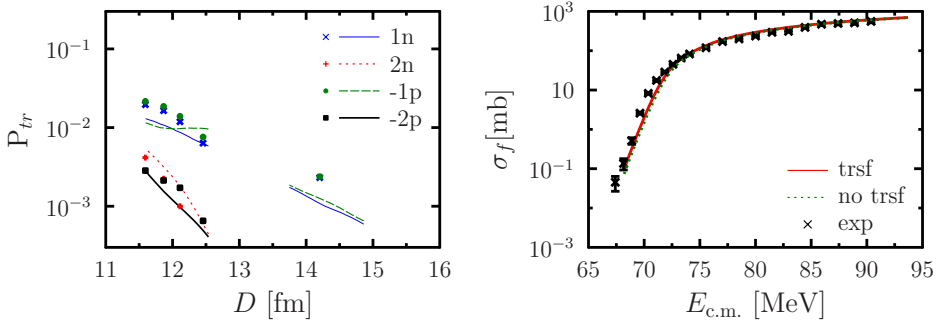}}
  \caption{  Left: The transfer probabilities obtained experimentally (marker) and with the 
coupled-channels method (lines) for the $^{40}$Ca+$^{58}$Ni reaction. Right: Corresponding fusion cross sections obtained experimentally and with the coupled-channels method with and without taking into account the couplings to the transfer channels (lines).   } \label{fig:prob_cross_58}
\end{figure}

\begin{figure}[!h]
  \centering \includegraphics[width=  \linewidth]{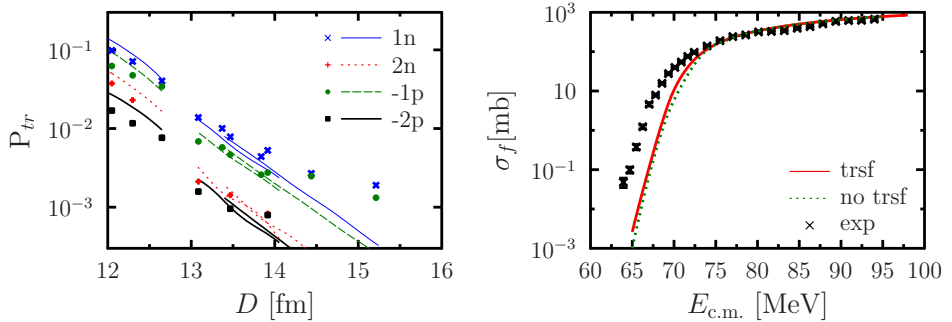}
  \caption{  Same as fig.  \ref{fig:prob_cross_58}  for the $^{40}$Ca+$^{64}$Ni reaction. } \label{fig:prob_cross_64}
\end{figure}

The transfer probabilities are computed as the ratio of the transfer cross sections to the elastic plus inelastic cross sections. The coupling parameters have been adjusted in order to reproduce the experimental transfer results as shown on the left part of  figs. \ref{fig:prob_cross_58} and \ref{fig:prob_cross_64}. In these figures, the transfer probabilities are shown as a function of the distance of closest approach $D$ based on the Rutherford trajectory. As expected from the Q-value, the transfer probabilities are by one order of magnitude less for the $^{40}$Ca+$^{58}$Ni system than for the $^{40}$Ca+$^{64}$Ni system. In the former system, the coupled-channels reproduces well the fusion cross sections because the transfer effect is too small to enhance the fusion cross sections. 

For the  $^{40}$Ca+$^{64}$Ni fusion reaction, the coupled-channels method predicts a small effect of the transfer. In contrast, experimentally the fusion cross sections are enhanced at energies below the barrier. Note that we can also invert the procedure, that is, by fitting the transfer coupling in order to reproduce the fusion cross sections and then comparing the transfer probabilities to the experimental data. In that case, the fusion cross sections will be well reproduced but the predicted transfer probabilities would be overestimated by about a factor of 10. In either way, it does not seem possible with this approach to reproduce simultaneously the transfer probabilities and the sub-barrier enhancement of fusion cross sections.


In conclusion, this analysis has shown a possible limitation of the present coupled-channels method. One possibility is the coupling scheme used in refs. \cite{Sca15, Esb16, Esb89}, that assumes that couplings to collective states are identical before and after the transfer. That is, the collective excitations and the transfer channels are treated as independent degrees of freedom.  More complex models may be necessary in order to reproduce the enhancement of fusion cross sections with the coupling strengths which reproduce 
the experimental transfer probabilities.

\section*{ACKNOWLEDGMENTS}
 G.S. acknowledges the Japan Society for the Promotion of Science
 for the JSPS postdoctoral fellowship for foreign researchers.
 This work was supported by Grant-in-Aid for JSPS Fellows No. 14F04769.



%
\end{document}